\documentclass[aps,prl,twocolumn,showpacs,preprintnumbers,amsmath,amssymb,superscriptaddress]{revtex4}
\usepackage{graphicx}
\usepackage{color}
\usepackage{ulem}
\begin{document}

\title{Extraction of many-body configurations from\\
nonlinear absorption in semiconductor quantum wells}

\author{R.P.~Smith}
\affiliation{JILA, National Institute of Standards and Technology and
 University of Colorado, Boulder, CO 80309-0440, USA}
\affiliation{Department of Physics, University of Colorado, Boulder, CO 80309-0390, USA}
\author{J.K.~Wahlstrand}
\affiliation{JILA, National Institute of Standards and Technology and
 University of Colorado, Boulder, CO 80309-0440, USA}
\author{A.C.~Funk}
\affiliation{JILA, National Institute of Standards and Technology and
 University of Colorado, Boulder, CO 80309-0440, USA}
 \author{R.P.~Mirin}
\affiliation{National Institute of Standards and Technology, Boulder, Colorado, 80305 USA}
\author{S.T.~Cundiff}
\email{cundiffs@jila.colorado.edu}
\affiliation{JILA, National Institute of Standards and Technology and
 University of Colorado, Boulder, CO 80309-0440, USA}
\affiliation{Department of Physics, University of Colorado, Boulder, CO 80309-0390, USA}
\author{J.T.~Steiner}
\affiliation{Department of Physics and Material Sciences Center,
Philipps-University Marburg, Renthof 5, D-35032 Marburg, Germany}
\author{M.~Schafer}
\affiliation{Department of Physics and Material Sciences Center,
Philipps-University Marburg, Renthof 5, D-35032 Marburg, Germany}
\author{M.~Kira}
\affiliation{Department of Physics and Material Sciences Center,
Philipps-University Marburg, Renthof 5, D-35032 Marburg, Germany}
\author{S.W.~Koch}
\affiliation{Department of Physics and Material Sciences Center,
Philipps-University Marburg, Renthof 5, D-35032 Marburg, Germany}

\date{\today}

\begin{abstract}
Detailed electronic many-body configurations are extracted from quantitatively measured time-resolved nonlinear absorption spectra of resonantly excited GaAs quantum wells. The microscopic theory assigns the observed spectral changes to a unique mixture of electron-hole plasma, exciton, and polarization effects. Strong transient gain is observed only under co-circular pump-probe conditions and is attributed to the transfer of pump-induced coherences to the probe.
\end{abstract}
\pacs{03.65.Wj,42.50.-p}

\maketitle

\newcommand{\be}{\begin{equation}}
\newcommand{\ee}{\end{equation}}
\newcommand{\bea}{\begin{eqnarray}}
\newcommand{\eea}{\end{eqnarray}}
\newcommand{\eps}{\varepsilon}
\newcommand{\ev}[1]{\langle#1\rangle}
\newcommand{\ddt}{\frac{\partial}{\partial t}}
\newcommand{\ihddt}{i\hbar\frac{\partial}{\partial t}}
\newcommand{\mcal}[1]{{\mathcal{#1}}}
\newcommand{\mrm}[1]{{\mathrm{#1}}}
\newcommand{\drm}{{\mathrm{d}}}
\newcommand{\e}{\mathrm{e}}
\renewcommand{\matrix}[1]{\mathbf{#1}}
\renewcommand{\vec}[1]{\mathbf{#1}}
\newcommand{\ci}[1]{\mathbf{#1}}
\newcommand{\stumm}{\bullet}
\newcommand{\ssum}[1]{\left[\left[#1\right]\right]}
\newcommand{\csum}[2]{\left|\left|#2\right|\right|_{#1}}
\newcommand{\widebar}[1]{\overline{#1}}


Precise extraction of a system's quantum state provides complete characterization of its physical properties. This ultimate level of characterization can only be realized in very few cases. Examples include determining the Wigner function of a single light mode using various quantum-state reconstruction techniques \cite{Smithey:93,Lvovsky:01,Bertet:02,Ourjoumtsev:07} or imaging molecular orbitals using attosecond techniques \cite{Kienberger:04,Itatani:04,Torres:07}. In contrast, the direct extraction of the full many-body configuration in solids seems inconceivable due to the overwhelmingly large number of degrees of freedom involved. Optically excited direct-gap semiconductors are an ideal candidate for determining the many-body configurations because the nonlinear optical properties of the excitonic absorption depend strongly and uniquely on the particular many-body state.

Resonant or nonresonant optical excitation of direct-gap semiconductors induces an optical polarization, i.e.,~a transition amplitude that can be converted into electron-hole (e-h) pair excitations and possibly into Coulomb-bound excitons \cite{Kira:04}. Thus, the actual many-body state contains a mix of polarization, e-h plasma, a fraction of true excitons, and higher-order correlations. Pump-probe and four-wave mixing measurements show that the excitonic resonances in the optical spectra depend on the excitation level \cite{Haug:09,Chemla:01,Koch:06}. Typically, excitonic resonances show pronounced nonlinear broadening \cite{Wang:93}, energetic shifts \cite{Peyghambarian:84}, and a transition from absorption to gain \cite{Chow:97} at high carrier density. Establishing a mapping between the nonlinear properties of the excitonic absorption and the underlying many-body state is challenging because a multitude of effects contribute to the nonlinearities. Noteworthy are Fermionic Pauli blocking, screening of the Coulomb interaction \cite{Haug:09}, carrier-polarization scattering \cite{Wang:93,Hu:94,Jahnke:96,Manzke:98,Shacklette:02}, polarization-polarization scattering \cite{Lygnes:97}, and exciton-exciton scattering \cite{Peyghambarian:84}.

In this Letter, we report {\it quantitative} measurements of quantum-well (QW) absorption spectra $\alpha_{\rm QW}$ after ultrafast resonant laser excitation. Furthermore, we use a fully microscopic, fit-parameter-free approach \cite{Kira:06} to compute a series of theoretical spectra $\alpha_{\rm th}$ resulting from a set of many-body states. By matching $\alpha_{\rm th}$ with $\alpha_{\rm QW}$, we extract the maximum-likelihood (ML) many-body configuration. Our results show that the polarization, density, and exciton population configurations can be accurately identified by relating specific nonlinear aspects of $\alpha_{\rm QW}$ to a particular many-body configuration. Such an accurate extraction can only be achieved by using {\it quantitative measurements}, as opposed to normalized measurements as is typically done.

We study a sample containing ten 10\,nm GaAs QWs separated by 10~nm thick ${\rm Al_{0.3}Ga_{0.7}As }$ barriers. At 4\,K, the QW heavy-hole $1s$ exciton resonance is at $E_{1s}= 1.546$\,eV and is well-separated from all other absorption resonances. We carefully measure the linear transmission and reflection spectra of the unexcited sample using white light. The linear results are analyzed with a transfer-matrix computation that includes all layers in the sample. We fine tune the refractive indices and thicknesses of the dielectric layers in the model to reach a full agreement between linear theory and measurements for the energy range 1.4--1.7\,eV. This procedure allows us to calibrate the nonlinear absorption experiments and determines all needed sample parameters, which are fixed in all our subsequent quantitative theory-experiment comparisons.

The nonlinear optical response is recorded in a pump-probe configuration using a mode-locked Ti:sapphire laser with a pump spectrum centered at $E_{1s}$ and a half-width half maximum (HWHM) bandwidth of 2.9\,meV to avoid exciting the e-h continuum or the light-hole resonance. The pump pulse is focused onto a 100\,$\mu$m spot. After a delay $\tau$, a low-intensity probe pulse arrives from a direction different from that of the pump in order to avoid direct pump-probe transfer. The weak probe is focused to a 20\,$\mu$m spot in the center of the pumped area to monitor only the spatially homogeneous part of the excitation. We record the probe transmission $T(\omega)$ and reflection $R(\omega)$ probabilities as functions of photon energy $\hbar \omega$ for a matrix of time delays, pump powers, and polarization configurations. Through simultaneous calibrated measurements of the incident, transmitted, and reflected probe powers, we determine the true absorption $  \alpha(\omega) = 1 - T(\omega)-R(\omega)$. For growth purposes, the sample includes bulk GaAs that produces a spectrally flat background absorption of 28\,\%. Thus, the QW absorption is $\alpha_{\rm QW}(\omega) \equiv  \alpha(\omega) -0.28$ for this sample.

\begin{figure}[!ht]
\center{\scalebox{0.38}{\includegraphics[angle=0]{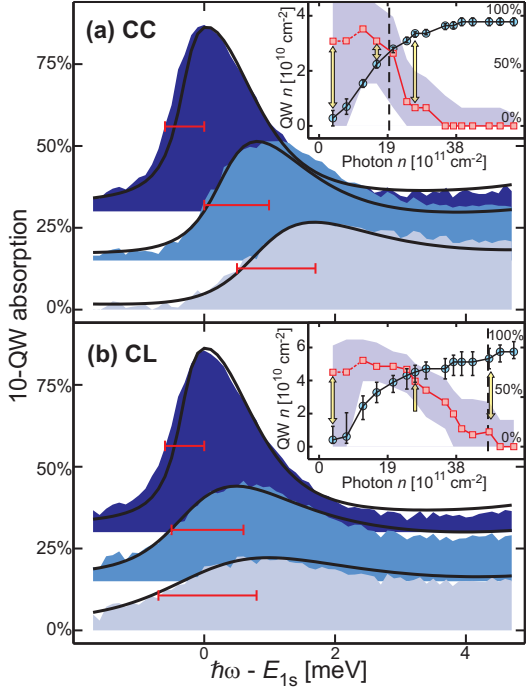}}}
\caption{
(Color online) Nonlinear QW absorption at large $\tau=13$\,ps delay. (a) The measured CC-configuration $\alpha_{\rm QW}$ (shaded areas) are shown for pump-photon densities $3.9$, $16$, and $27\times10^{11}\,{\rm cm}^{-2}$ and compared to microscopically computed spectra (solid lines) for densities $n=0.14$, $2.2$, and $3.4\times{}10^{10} {\rm cm}^{-2}$, from top to bottom. The spectra are vertically offset by 15\,\%. The horizontal lines indicate the HWHM of the respective exciton resonances.
(b) The corresponding CL comparison with $3.9$, $27$, and $48\times10^{11}\,{\rm cm}^{-2}$ pump-photon densities (shaded areas) and calculations for the densities $0.21$, $4.4$, and $5.0 \times 10^{10}\,{\rm cm}^{-2}$ (solid lines).
Insets: The deduced ML carrier densities (circles, left scale) and exciton fraction (squares, right scale) for each experiment. The arrows indicate the spectra presented in the respective figures; the dashed vertical lines mark saturation.
}
\label{fig:long}
\end{figure}

Figure \ref{fig:long} presents the measured $\alpha_{\rm QW}$ for three different pump powers. The probe and pump have (a) co-circular (CC) and (b) co-linear (CL) polarizations while the pump-probe delay is $\tau=13$~ps. In each case, the lowest-intensity $\alpha_{\rm QW}$ is in the linear regime and shows a clear $1s$ resonance, yielding $\alpha_{\rm QW} = 57\,\%$ with a HWHM $\gamma_{1s}=0.75$\,meV (horizontal bars). Roughly half of this width is estimated to result from radiative decay in the nontrivial dielectric structure \cite{Kira:06}. When the excitation power is increased, $\alpha_{\rm QW}$ experiences nonlinear changes. The $1s$ resonance broadens and becomes asymmetric, similarly to prior observations \cite{Wang:93,Jahnke:96,Shacklette:02}. We also see that the CC and CL configurations yield very different nonlinearities. Under CC conditions, we obtain an almost 2\,meV blue shift of the $1s$ resonance while the CL configuration only exhibits a smaller, less than 1\,meV, blue shift before saturation is reached. The nonlinear CL spectra are about 50\% broader than the CC spectra.

We investigate the nonlinear response using our microscopic theory \cite{Kira:06} to identify the relevant many-body state. More explicitly, we evaluate the dynamics of the microscopic, probe-generated, polarization $\delta p_{\bf k}$ at the crystal momentum ${\bf k}$. For each QW, we have the semiconductor Bloch equations \cite{Haug:09,Kira:06},
\begin{eqnarray}
  i \hbar \frac{\partial}{\partial t} \delta p_{\bf k}
  &=& \epsilon_{\bf k} \delta p_{\bf k}
     - (1 -f^e_{\bf k} -f^h_{\bf k} ) \; \delta \Omega_{\bf k}(t)
     +\Gamma_{\bf k}
\nonumber\\
  &&-P_{\bf k} \sum_{\lambda,{\bf k}'} \delta f^\lambda_{{\bf k}'}
    + (\delta f^e_{\bf k} + \delta f^h_{\bf k}) \; \Omega_{\bf k}
\label{eq:probe_SBE},
\end{eqnarray}
that contain the pump-generated electron (hole) distribution $f^{e(h)}_{\bf k}$, the polarization $P_{\bf k}$, the Coulomb-renormalized energy $\epsilon_{\bf k}$, and the renormalized Rabi energies, $\Omega_{\bf k} \equiv d \, E_{\rm pump} +\sum_{{\bf k}'} V_{{\bf k}'-{\bf k}} P_{{\bf k}'}$ and $\delta\Omega_{\bf k} \equiv d \, E_{\rm pro} +\sum_{{\bf k}'} V_{{\bf k}'-{\bf k}} \delta p_{{\bf k}'}$. The interactions involve the dipole-matrix element $d$, the electric field $E_{\rm pump}$ ($E_{\rm pro}$) for the pump (probe), and the Coulomb-matrix element $V_{\bf k}$. The probe polarization $\delta p_{\bf k}$  couples to two-particle Coulomb correlations $\Gamma_{\bf k}$ that produce screening of the Coulomb interaction, higher-order energy renormalizations, and Boltzmann-type scattering of $\delta p_{\bf k}$ from various quasiparticles and transition amplitudes \cite{Kira:06}. When the pump $P_{\bf k}$ is still present, the carrier densities are changed linearly by $\delta f^\lambda_{\bf k}$ with dynamics
\begin{eqnarray}
   \hbar \frac{\partial}{\partial t} \delta f^\lambda_{\bf k}
  = 2
     {\rm Im}
        \left[
           \delta\Omega^{\star}_{\bf k} \, P_{\bf k}
           +
           \Omega^{\star}_{\bf k} \, \delta p_{\bf k}
        \right]
     +r_{\bf k}
\label{eq:probe_f},
\end{eqnarray}
where $r_{\bf k}$ describes the scattering.

The probe response follows directly from the linear QW susceptibility $\chi(\omega) \equiv \frac{1}{\cal S} \sum_{\bf k} \frac{ \delta p_{\bf k}(\omega)}{E_{\rm pro} (\omega)}$ after we Fourier transform $E_{\rm pro}$ as well as $\delta p_{\bf k}$ and use the normalization area ${\cal S}$. The true QW absorption is then evaluated via a transfer-matrix computation that has both  $\chi(\omega)$  and the full dielectric structure of the sample as inputs. This procedure includes all of the Coulomb-induced nonlinearities discussed above for a given pump-generated many-body configuration. The mean-field quantities, i.e.~$f^{e(h)}_{\bf k}$ and $P_{\bf k}$, represent the simplest part of the excitation configuration, while bound exciton and biexciton amplitudes determine pair-wise correlations. The microscopic influence of all these quantities on $\chi(\omega)$ is systematically evaluated with the cluster-expansion approach \cite{Kira:06}. We find that $\alpha_{\rm th}$ is extremely sensitive to the pump-induced $f^{e(h)}_{\bf k}$, $P_{\bf k}$, and exciton populations.

As a major extension beyond our published quantum-wire analysis \cite{Kira:04,Kira:06}, we solve Eqs.~\eqref{eq:probe_SBE}--\eqref{eq:probe_f} for a realistic QW system and a large number of combinations of $f^{e(h)}_{\bf k}$, $P_{\bf k}$, and exciton populations. Under quasi-equilibrium conditions, we use $f^{e(h)}_{\bf k}$ in the form of a Fermi-Dirac distribution defined by the temperature and the carrier density $n=\frac{1}{\cal S} f^{e(h)}_{\bf k}$. Due to the resonant excitation configuration, we also include the polarization and the fraction $x$ of carriers bound into $1s$ excitons. Altogether, we numerically evaluate $\alpha_{\rm th}$ for more than $10^5$ different many-body configurations. To extract the ML configuration, we define the normalized deviation
\begin{eqnarray}
   \epsilon
   \equiv
   \frac{
       \int d\omega \;
       \left|
          \alpha_{\rm th}(\omega) - \alpha_{\rm QW}(\omega)
       \right|
    }
    {
       \int d\omega \;
       \left|
          \alpha_{\rm QW}(\omega)
       \right|
       }
\label{eq:deviation},
\end{eqnarray}
between the computed and measured QW absorption spectrum. We minimize $\epsilon$ via a multi-dimensional optimization, yielding the ML excitation configuration. We determine a confidence interval of the configurations yielding below $5\%$ variations in $\epsilon$.

Figure~\ref{fig:long} shows the $\alpha_{\rm th}$ (solid lines) of the extracted ML configuration that reproduces the respective experimental result (shaded areas). The inset gives information about the actual ML configuration as a function of the photon density within the pump. The circles indicate the ML carrier density $n$ with error bars marking the confidence interval for each different experiment separately. The squares show the ML fraction of excitons while the shaded area indicates the related confidence interval. Since the pump-generated polarization has completely decayed for long delay times, there is no $P_{\bf k}$ present in the ML configurations. We also find that the e-h density first grows linearly as a function of excitation power and then saturates. This saturation is simply explained by the ionization of the $1s$-exciton resonance for elevated densities, which strongly reduces the ability of the pump to generate more carriers. Saturation occurs close to that ML configuration where the phase-space filling factor $1-f^e_0-f^h_0$ becomes zero (dashed vertical line). For the cases studied, the best-fit electron and hole temperatures are 38\,K and 11\,K, respectively. At the same time, the exciton fraction is initially high due to the almost 100\% efficiency in polarization-to-population conversion. Onset of ionization causes the exciton fraction to rapidly decrease as a function of excitation density.

The pronounced differences between the CL and CC experiments are related to the fact that the CC case excites carriers only into one spin state, while under CL conditions, one populates both spin states. Since the Coulomb-scattering of $\delta p_{\bf k}$ with the e-h plasma is spin sensitive, the CL and CC cases lead to different Coulomb-induced shifts of the $1s$ resonance.
The fact that the shifts persist to saturation, where no excitons exist, verifies that the e-h plasma and not the exciton population is mostly responsible for the blue shift of the $1s$ resonance. The enhanced broadening of the CL relative to the CC case is explained by the fact that there are twice as many scattering partners for $\delta p_{\bf k}$ in the CL case due to the e-h occupations in both spin states.

\begin{figure}[!ht]
\center{\scalebox{0.38}{\includegraphics[angle=0]{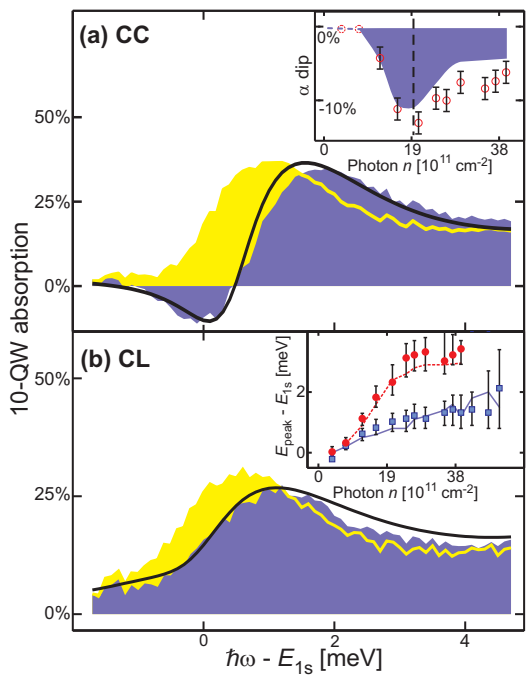}}}
\caption{
(Color online) Nonlinear QW absorption for the short $\tau=3$~ps delay.
(a) CC conditions with photon density $16\times10^{11}\,{\rm cm}^{-2}$:
the measured $\alpha_{\rm QW}$ (dark shaded), the computed
 $\alpha_{\rm th}$ (solid line), and the long-time delay $\alpha_{\rm QW}$ (light shaded). Inset: the measured (open circles) and computed (shaded area) absorption dip (gain maximum) as a function of pump-photon density. The dashed vertical line marks saturation.
(b) Same as (a), but for the CL configuration with the photon density $27\times10^{11}\,{\rm cm}^{-2}$.
Inset: the absorption peak shift for CL (squares) and CC (circles) measurements compared with theory (lines).
}
\label{fig:short}
\end{figure}

As a further test of our ML extraction procedure, we next consider short pump-probe delay conditions where coherent transients \cite{Sieh:99} as well as excitonic gain \cite{Mysyrowicz:86} have been observed previously. Figure \ref{fig:short} shows the measured $\alpha_{\rm QW}$ (dark area) and the extracted ML $\alpha_{\rm th}$ (solid line) just after the excitation ($\tau = 3\,$ps) for (a) the CC and (b) the CL excitation. The short delay $\alpha_{\rm QW}$ and $\alpha_{\rm th}$ not only agree but they are also distinctly different from the long delay $\alpha_{\rm QW}$ (light shaded area). In particular, the CC case yields a 15\% gain feature (negative absorption) just below the $1s$ resonance. As a main difference to the long delay investigations, the short-delay ML configuration contains a significant portion of pump-generated $P_{\bf k}$. Since the presence of the induced QW polarization is needed to produce gain, the gain is transient and can be attributed to coherent polarization transfer between pump and probe.

The transient gain delicately depends on the excitation conditions. It is not seen for the CL case due to the above mentioned enhanced dephasing of $\delta p_{\bf k}$. For CC excitation, we obtain the strongest transient gain for elevated excitation densities close to the $1s$-saturation. To obtain analytic insight on how $\delta f^\lambda_{\bf k}$ influences the coherent transients, we also solve Eqs.~\eqref{eq:probe_SBE}--\eqref{eq:probe_f} for the case where scattering is omitted. We find that $\delta f^\lambda_{\bf k}$ is proportional to $\left(1 - f^e_{\bf k} - f^h_{\bf k} \right)^{-1}$ showing that the coherent transients become increasingly strong near the $1s$-exciton saturation because the phase-space filling factor approaches zero.
To verify this, we plot in the inset to Fig.~\ref{fig:short}a the measured maximum values of $\alpha_{\rm QW}$ gain (open circles) as a function of pump density. We clearly see that the transient gain reaches its maximum value close to the $1s$-ionization threshold (saturation), indicated by the vertical dashed line. The inset to Fig.~\ref{fig:short}b presents the energetic position of the measured CL (squares) and CC (circles) $\alpha_{\rm QW}$, while the lines are from the ML calculations. We find that the pump-generated polarization also produces an additional blue shift for both CL and CC excitations. In particular, the short delays produce blue shifts up to 2\,meV (CL) and 4\,meV (CC), which are roughly a factor of two larger than for those at long delays.

\begin{figure}[!ht]
\center{\scalebox{0.38}{\includegraphics[angle=0]{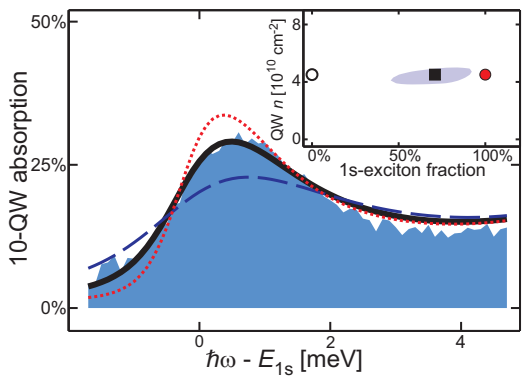}}}
\caption{
(Color online) Influence of exciton populations on nonlinear $\alpha_{\rm QW}$. The long-delay measured $\alpha_{\rm QW}$ (shaded area) for $16\times10^{11}\,{\rm cm}^{-2}$ pump is compared with the ML result (solid line) as well as computations having 0\% (dashed line) and 100\% (dotted line) excitons.
Inset: The ML configuration (square) and the 5\% confidence interval (shaded area) plotted in $(n,x)$ space. The 0\% (open circle) and 100\% (filled circle) represent exciton configurations used for the dashed and dotted lines, respectively.
}
\label{fig:max-Likelihood}
\end{figure}

The ML analysis can also be used to quantitatively determine the role of excitons. The inset to Fig.~\ref{fig:max-Likelihood} shows the 5\% $\epsilon$ confidence interval (shaded area) in carrier density and exciton fraction when the intermediate-intensity CL experiment of Fig.~\ref{fig:long}b is analyzed. The overall difference between theory and experiment is minimized around a single $(n,x)$ configuration (square). Figure \ref{fig:max-Likelihood} shows the corresponding experimental $\alpha_{\rm QW}$ spectrum (shaded area) together with the ML result (solid line) as well as cases where the exciton fraction is reduced to $x=0$\% (dashed line) or raised to $x=100$\% (dotted line). Increasing the exciton fraction leads to a reduction of the excitation-induced broadening and blue shifting of the $1s$-resonance.  Intuitively, these effects follow because the scattering of the probe induced $\delta p_{\bf k}$ becomes less likely when electrons and holes are bound into charge-neutral excitons ($x=100\%$) than when they remain as an ionized plasma ($x=0\%$).
It is also interesting to see that formation of excitons ($x=100\%$) yields only a small energy renormalization that does not produce an additional blue shift, in contrast to polarization studied in Fig.~\ref{fig:short}.

In conclusion, our detailed comparison between quantitative experiments and theory shows that the absorptive nonlinearities in $\alpha_{\rm QW}$ depend so sensitively on the many-body configuration that we can extract the maximum likelihood many-body excitation configuration with great confidence. The analysis identifies the role of coherent polarization, exciton and electron-hole plasma contributions. For the short-time co-circular pump-probe configuration, pronounced transient gain is observed with a strength controlled by the pump.

The work at JILA was supported by NIST and the NSF.
The work in Marburg was supported by the Deutsche Forschungsgemeinschaft.

\end{document}